# Reinforcing Iron Metal Matrix Composite by Multi-Wall Carbon Nanotube: A Combined Theoretical and Computational Approach

Raashiq Ishraaq[1,*], Mahmudur Rashid[1], Shahriar Muhammad Nahid[2]

[1]Department of Mechanical Engineering, Bangladesh University of Engineering and Technology, Dhaka 1000, Bangladesh
[2]Department of Mechanical Science and Engineering, University of Illinois Urbana Champaign, Urbana, IL 61801, United States

*Corresponding author:ishraaqraashiq@gmail.com

## Abstract

Carbon nanotube (CNT) reinforced metal matrix composites have been the focus of researchers due to their high load-bearing capacity. Among single and multi-wall carbon nanotubes (MWCNT), the latter is preferred by manufacturers and engineers for making composites due to their economic feasibility of synthesizing. However, the effect of layer numbers along with other parameters of the reinforcing MWCNT must be understood before its industrial application. In this article, we developed a novel theoretical approach for predicting the variation of strength and stiffness of MWCNT reinforced iron composites (MWCNT-Fe) with the layer number of reinforcing MWCNT and validated the prediction with a series of Molecular dynamics (MD) simulation. Our analysis revealed that for every addition of two extra layers, the strength and stiffness of the composite increase 9.8% and 7.2% respectively up to eight layered MWCNT and then becomes saturated. We also employed MD simulations for investigating the effect of grain boundary on the failure mechanism of CNT reinforced iron composites in contrast to previous studies. Our investigations revealed that instead of the matrix-fiber interface, the failure was initiated from the grain boundary and merges with the interface. The results in this study will not only help engineers and manufacturers choose optimal layered MWCNT for synthesizing composite for a specific application but also provide scientists a new method to model composites for predicting desired properties.

# 1. INTRODUCTION:

After the discovery of CNT by Ijima [1], material scientists and engineers tried to incorporate its outstanding mechanical [2,3], thermal [4], and electrical [5] properties to create a new generation of materials [6–9]. Among these materials, research activities related to CNT reinforced composites have gained popularity among the scientific community for their high mechanical strength and stiffness [2,10,11]. Previously, researchers were focused on developing and studying various types of CNT reinforced polymer composites [12], [13]. However, in the present decade, CNT reinforced metal matrix composites (CNT-MMC) have captured the interest of metallurgists due to their enormous potential in space, automobile, and high precision device industries [13–16]. Mixing CNT with metals enhances its mechanical properties significantly reducing the amount of metal required for strength applications. In consequence, vehicles and aircrafts manufactured with metal matrix composites (MMC) have better mileage and reduced handling costs.

In recent times, many researchers synthesized samples of CNT-MMC and performed various tests to determine their mechanical properties. Ao et al. reinforced Zn with Cu coated MWCNT by the combined use of electroless deposition (ED), spark plasma sintering (SPS), and hot-rolling techniques. They reported that composites reinforced with Cu coated CNT demonstrated enhanced mechanical properties compared to uncoated CNT and reinforcement with 0.3 vol% Cu coated CNT showed the highest strength (280 MPa) [17]. The enhanced load transfer effect for Cu coating and fine grain strengthening were the two mechanisms for the composite's increased strength. In the study conducted by Wang et al. [18], CNT-Cu was made by ED and SPS process. They found that embedding CNT on Cu enhanced the mechanical properties of the composite significantly and the composite containing 0.5% vol CNT showed maximum Vickers hardness number ( 1.3 GPa ) and yield strength ( 142.2 GPa ). CNT-Mg composites having different wt% CNT were fabricated by melt dispersion and hot extrusion by Goh et al. Their investigations revealed that the yield strength, tensile strength, and ductility increased up to 1.3 wt% CNT and then decreased due to high amounts of porosity in the Mg matrix [19].

In general, the inclusion of CNT creates a new type of anomaly in the lattice configuration of the metal. Therefore, atomistic studies are required to know the behavior of the composites under different environmental conditions and microscopic structures prior to their manufacturing and industrial usage. Due to the difficulty of accurately conducting experiments at the nanoscale, computational approaches are quite efficient and popular. In recent years, MD has emerged as a proficient computational method for studying composites at an atomistic level. For example, Song & Zha investigated the difference in the enhancement of mechanical properties of gold for embedding long CNT and short CNT [20]. They reported that reinforcing gold with long CNT increased its stiffness but short CNT decreased its yield stress and strain due to acting as a void inside the composite. An MD study by Sharma et al. compared the mechanical reinforcing effect of CNT and single-layered graphene on the Nickel matrix [21]. Song and Zha [22] studied the effect of Ni coating on CNT in an aluminum matrix. Their study revealed that Ni coating on CNT enhances the stiffness of the composite compared to the uncoated CNT due to enhancing load transfer from matrix to fiber. Cong and Lee [23] used MD to explore the effect of fiber size and volume fraction of boron nitride nanotubes (BNNT) reinforced aluminum composite. They found that embedding (4,4), (6,6), (8,8) and (10,10) BNNT on aluminum increased its stiffness by 15.04%, 22.61%, 30.53% and 41.75% respectively.

Though iron is the most used metal on earth, very few experimental and computational studies have been done related to CNT nanotube-reinforced iron composite (CNT-Fe). Recently, Suh and Bae synthesized MWCNT-Fe composites using hot pressing and sintering techniques and reported that the composite containing 2% and 4% MWCNT respectively had 25% and 45% higher compressive strength than pure iron [24]. Dong et al. [25] investigated the effect of mixing CNT on the Fe-Cu binder of diamond cutting tools. Parswajinan et al. [26] studied the mechanical reinforcement effect of MWCNT on iron by manufacturing the composite using the powder metallurgy technique. They reported that for adding only 0.67 wt% CNT, the tensile and compressive strength of iron increased by 17.12% and 22.03% respectively. However, these experimental studies focused only on measuring the effect of different manufacturing parameters (CNT radius, manufacturing techniques, CNT percentage, etc.) on the mechanical properties of the composite and did not provide any theoretical framework. In our previous study [27], we developed a theoretical model of CNT-Fe composite to predict the trend of the composite's mechanical properties with CNT radius, which was verified by MD simulations. Our investigation revealed that CNT having a radius of 2.4 Å gives the highest mechanical strength for a given matrix volume fraction.

However, to this date, many important factors affecting CNT-Fe composite's mechanical performance are still uninvestigated. From the experiment, it is also confirmed that iron composites reinforced by multiwall carbon nanotube (MWCNT) have higher mechanical properties than the ones reinforced by single-wall carbon nanotube (SWCNT) [28]. Nevertheless, how the strength and stiffness of the composite vary with the layer number of MWCNT is still unclear. Results obtained from experimental studies for CNT MMC remain unreliable for investigating such properties due to the huge variation of data among literature, which arises because of the poor dispersion of CNT in metal and many other factors [29]. There is also a lack of theoretical studies to qualitatively predict the properties of the composites with the changes in manufacturing parameters, which could have helped researchers to validate their experimental data. It is also a known fact that real composites have grains in their matrix instead of a single crystal [30]. Investigating the interplay between grain boundaries and the matrix-fiber interface can improve our understanding of the failure process and pave the way for further improving its mechanical properties. Yet, no studies were done for examining how the grain boundaries interact with the matrix-fiber interface and affect the failure mechanism.

In this article, in order to address the above-mentioned problems, we developed a novel theoretical model with the basic knowledge of composite materials and continuum mechanics and predicted the trend of mechanical properties for increasing layers of MWCNT. Then we validated our predicted trend with MD simulations for composites having a single and poly grain matrix. We also investigated the failure mechanism of poly grain CNT-Fe composites. Our atomistic inspection provided a detailed insight into the composite's failure mechanism and also revealed how the composite's strength and stiffness will vary with grain sizes below the critical grain size. Our study will be helpful to farther suitable manufacturing processes to synthesize ultra-fine or nano-grained CNT-Fe MMC for making high precision nanodevices [31][32].

This paper is organized in the following way. In section 2, a detailed description of the MD simulation model, potential functions, and the simulation methods are given. Section 3 provides the results and observations and the final section summarizes the findings of this study and suggests future prospects for further investigation.

## 2. METHODOLOGY:

For MD simulations, we constructed four rectangular atomistic models of the representative volume element (RVE) of CNT-Fe composites, which are reinforced with two, four, eight and thirteen-layered MWCNT respectively. The axis of the CNTs was parallel with the z-axis and the length of all RVE models along that direction was 54.3 Å. The length of each model along the x and y direction was equal, which resulted in the RVE models having a square x-y cross-section. To keep the matrix volume fraction of each RVE equal (0.88), the dimensions along the x, y-axis were changed depending on the radius of the outermost layer of the MWCNT. The matrix volume fraction was held constant so that its effect does not interfere while investigating the influence of MWCNT layer numbers on the mechanical properties of the composite. The relation between the chirality of two consecutive CNT layers was maintained by following the equation $m = n + 5$, where $m$ is the chirality of the outer and n is the chirality of the inner CNT. This rule was followed to keep the graphitic distance (3.4 Å) between two CNT layers as suggested by Zhang et al. [33]. In order to investigate the viability of the predicted trend on a poly-grained matrix, two sets of RVE models having an average grain size of 5.31 nm$^3$ and 15.93 nm$^3$ were created. Each set contains four RVE models of CNT-Fe composites reinforced with MWCNT having two, four, eight and sixteen layers separately. Fig.1(a) shows one such model.

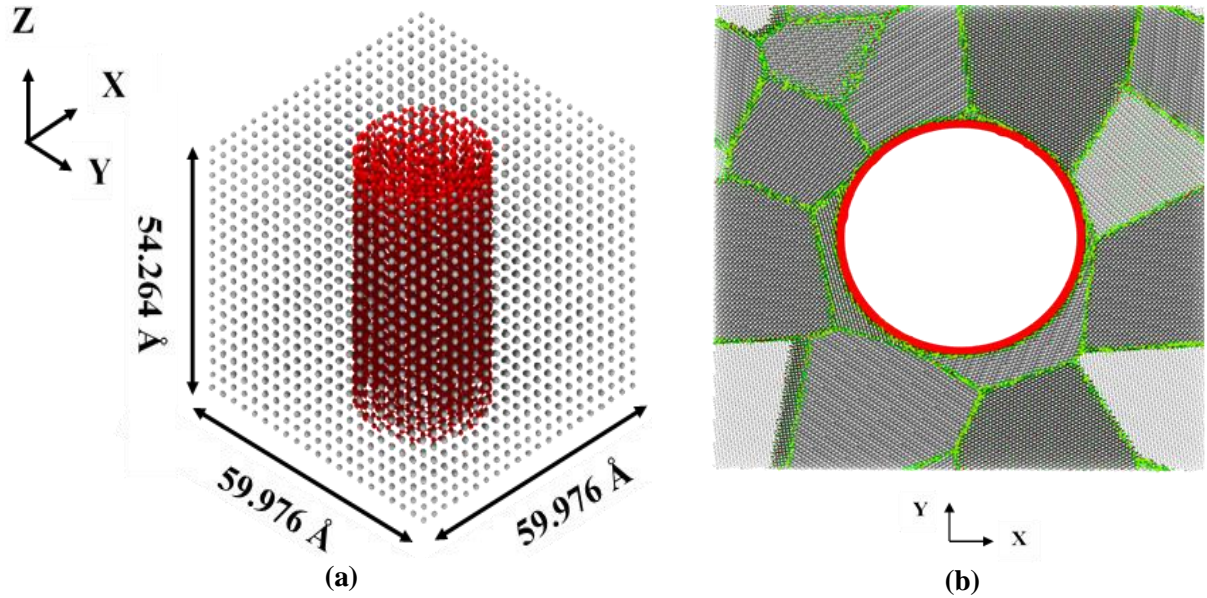

**Figure. 1:** RVE model of CNT-Fe composite reinforced by four layered MWCNT and having 0.88 matrix volume fraction. **(a)** shows the dimensions and overall view of the modeled RVE and **(b)** shows the top view of the RVE used to study the effect of grain size and failure mechanism of polygrain CNT-Fe composite. The green atoms in **(b)** shows the grain boundary.

For investigating the failure mechanism and the effect of grain size, we modeled three RVE of SWCNT reinforced Fe composite each having length of 291.31 Å,291.31 Å, and 54.26 Å along the x,y,z directions respectively. The size of these three RVE was bigger to capture the interactions between the grain boundary and the matrix-fiber interface in detail. Fig.1(b) depicts the top view of one of the RVE models.

Large-scale Atomic/Molecular Massively Parallel Simulator (LAMMPS) [34] was used to make the iron matrix and to perform simulation while visual molecular dynamics (VMD) [35] was used to create the CNT. Atomsk [36] software was used to build the RVE models having a poly-grain matrix when studying the effect of grain sizes OVITO [37] was used for visualization.

Three types of potentials are used to build these models: (1) Embedded atom method (EAM) (2) AIREBO and (3) Lennard Jones.

EAM potentials describe the metallic bonds with high accuracy [38]. Therefore, it is used to model the Fe-Fe interaction in this study. The parameters of the EAM potential employed in this study was developed by Auckland et al. [39]. Mathematically, EAM potentials can be described as below:

$$E_i = F_\alpha \left( \sum_{j \neq i} \rho_\beta(r_{ij}) \right) + \frac{1}{2} \sum_{j \neq i} \phi_{\alpha\beta}(r_{ij}) \tag{1}$$

In the above equation, $F$ is the embedding energy, which is also a function of electron density ($\rho$). The pair interactions are captured by the pair potential function, $\phi$. Both $\rho$ and $\phi$ are functions of $r_{ij}$, which represents the interatomic distance between atom $i$ and $j$. Subscripts $\alpha$ and $\beta$ used in Eq.(1) denote the element types of atom $i$ and $j$ respectively.

AIREBO [39] potential was used to describe the C-C atomic interaction. The following equation describes the AIREBO potential:

$$E = \sum_i \sum_{i \neq j} [E_{ij}^{REBO} + E_{ij}^{LJ} + \sum_{k \neq i} \sum_{l \neq i,j,k} E_{i,j,k,l}^{TORSION}] \tag{2}$$

The terms $E_{ij}^{REBO}, E_{ij}^{LJ}$ and $E_{i,j,k,l}^{TORSION}$ in the above equation represents the REBO, Lennard Jones, and Torsional energy of the system respectively. The REBO, Lennard-jones and torsional terms of the system separately describe the covalent interactions, the nonbonded interactions, and the dihedral angle effect. REBO cutoff parameter was set to 2.0 for accurately modeling the fracture process of the CNT [40].

Only the Vander wall's interaction was considered responsible for the adhesion between Fe and C atoms at the matrix-fiber interface [41]. This mode of interaction is also existent between the C atoms of different CNT layers [42]. To model the Vander wall's force, 6-12 Lennard jones potential was used [43], which is expressed by Eq.(3).

$$E = 4\epsilon\left[\left(\frac{\sigma}{r}\right)^{12} - \left(\frac{\sigma}{r}\right)^6\right] \tag{3}$$

In Eq.(3), *r* is the interatomic distance between two atoms, $\sigma$ denotes the interatomic distance between two particles where the potential energy is zero and $\epsilon$ is the potential well depth. The $\sigma$ and $\epsilon$ values between two different types of atoms were found using the Lorentz-Bertholet rule [41] as described by Eqs 4.1 and 4.2. The parameters for all pairs of atomic pair interactions were obtained from previous studies [27], [44]. The value of the equilibrium distance (h) between the Carbon and Iron atoms along the matrix-fiber interface was according to, h=0.8584$\sigma$ as suggested by Jiang et al.[45] and used by Srivastava et al. [46].

$$\sigma = \left(\frac{\sigma_{11} + \sigma_{22}}{2}\right) \tag{4.1}$$

$$\epsilon = \sqrt{(\epsilon_{11}\epsilon_{22})} \tag{4.2}$$

At first, the conjugate gradient method was used to minimize the energy of the initial structure, which eliminates any overlapping of atoms. Then NVT and NPT equilibration methods were consecutively used for 100000-time steps each to stabilize the model at 300K and 1 atm. Periodic Boundary Condition (PBC) was maintained in all directions. Simulation of uniaxial tension was performed at a strain rate of 0.001 ps$^{-1}$ on these models and the stress-strain graph was generated, from which the ultimate strength and stiffness were obtained. To calculate Young's modulus from the stress-strain graph, the least square regression method was used in the linear elastic region.

The accuracy of the computational model and potential parameters for giving qualitative information was verified with experimental data in our previous study [27].

## 3. RESULTS AND DISCUSSION:

### 3.1 Effect of layer numbers of the reinforcing MWCNT:

For understanding the effect of the number of layers of MWCNT on the mechanical properties of CNT-Fe, a theoretical prediction is needed to ensure the validity of any experimental results due to the large variation in data from different studies. However, currently, there are no mathematical models for this purpose.

Therefore, a model of the RVE is developed following the continuum approach and using that model a function is created to theoretically predict the variation pattern of mechanical properties with the MWCNT layer number. The predicted pattern was then verified by MD simulations. The process for developing the function is explained systematically.

$$\sigma_c = \sigma_f V_f + \sigma_m V_m \qquad (5)$$

$$E_c = E_f V_f + E_m V_m \qquad (6)$$

In Eqs. (5) and (6); $\sigma_m, \sigma_f, E_f, E_m, V_f$ and $V_m$ denotes the tensile strength of the matrix, the tensile strength of fiber, Young's modulus of fiber, Young's modulus of the matrix, fiber volume fraction, and matrix volume fraction respectively. In these two equations, the ultimate strength and stiffness of the composite are expressed in terms of the strength and stiffness of its constituent elements by using the mixture rule.

A composite material is considered to reach its failure strength when any of its primary constituent material (matrix or fiber) fails. In this case, the iron matrix fails first because its failure strain is much lower than CNT. Considering this fact, Eq.(4) can be written as Eq.(7) where $\epsilon_{m,ut}$ denotes the failure strain of the matrix.

$$\sigma_{c,ut} = E_f \, \epsilon_{m,ut} \, V_f + E_m \, \epsilon_{m,ut} \, V_m \qquad (7)$$

From Eqs. (6) and (7) It can be observed that $E_f, \epsilon_{m,ut}, E_m, \sigma_f$ and $\sigma_m$ are constant since they are material properties. In this study, we are observing only the effect of the layer number of MWCNT. Therefore, the matrix volume fraction was kept constant as previously explained. So, Eqs. (6) and (7) can be expressed as Eqs. (8) and (9).

$$E_c = (constant)\, V_f + (constant) \qquad (8)$$

$$\sigma_{c,ut} = (constant)\, V_f + (constant) \qquad (9)$$

From the above two equations, it can be concluded that for a constant matrix volume fraction ($V_m$), the composite's stiffness and strength will depend only on the fiber volume fraction ($V_f$). So, if varying the number of layers of MWCNT changes $V_f$ then $E_c$ and $\sigma_{c,ut}$ will also be changed. However, the variation pattern of $V_f$, $\sigma_{c,ut}$ and $E_c$ with the number of layers of MWCNT will be similar because these quantities are proportional to each other according to Eqs. (8) and (9).

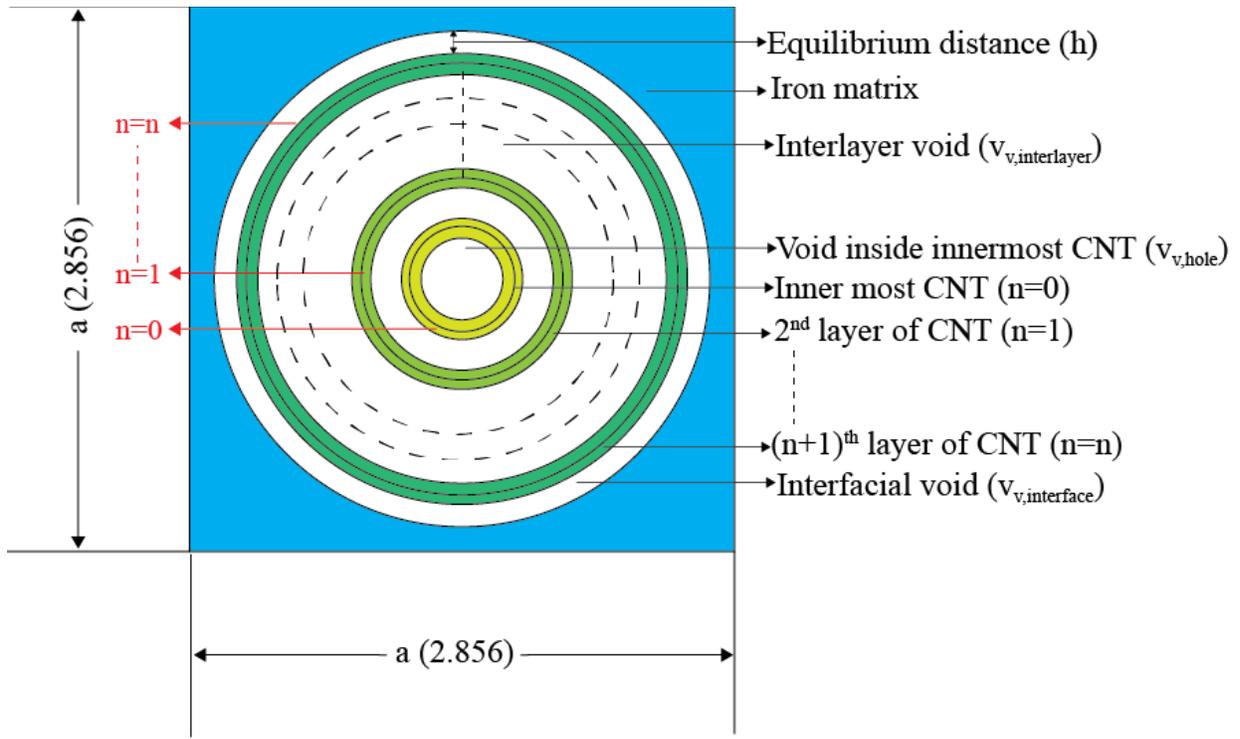

**Figure. 2**: The schematic diagram of the theoretical RVE model.(the figure is not drawn to scale)

Here, at first, we created an RVE model following the continuum approach and theoretically found the variation pattern of $V_f$ with the MWCNT layer numbers. Then we obtained $\sigma_{c,ut}$ and $E_c$ values from the simulation, which was used to verify the theoretical model.

The theoretical model of the RVE contains an ($n$ +1) layered MWCNT. The value of n ranges from zero for an SWCNT to any positive integer $n$, for a ($n$ +1) layered MWCNT. Fig.(2) shows the schematic diagram of the RVE model, where only the innermost CNT, the second layer, and the outermost layer of the MWCNT are shown for visual clarity. The radius of each CNT layer is considered to be the distance from the center to its midsection.

For a composite, the summation of $V_m$, $V_f$, and the void volume fraction ($V_v$) is equal to unity as given by Eq.(10). For a constant matrix volume fraction, the summation of fiber and void volume fraction also becomes a constant. Therefore Eq.(10) can be written as Eq.(11).

$$V_f + V_m + V_v = 1 \quad (10)$$

$$V_f + V_v = C \quad (11)$$

From Eq.(11), it can be observed that, the value $V_f$ and $V_v$ are also inversely proportional and interdependent. We defined $Z$ as described by Eq.(12), which is a function of the extra number of layers ($n$) of the MWCNT other than the innermost CNT. The output of the function gives the difference between $V_f$ and $V_v$ for a given $n$. Examining Eq.(12) it can be concluded that, the variation pattern of $V_f$ with $n$ is similar to the variation pattern of $Z$ because these two quantities are proportional. Therefore, finding the pattern of $Z$ will give the variation pattern of $V_f$, $\sigma_{c,ut}$ and $E_c$ with respect to $n$. To find the mathematical expression of $Z$, both $V_f$ and $V_v$ should be expressed only in terms of $n$ only.

$$Z(n) = V_f - V_v \quad (12)$$

From Fig. (3), it can be observed that the radius of the first and second CNT layer can be expressed in terms of CNT thickness ($t$), interlayer distance ($x$), and the radius of the innermost CNT ($r_0$) as shown by Eqs. (13) and (14). Following the same approach, the radius of the $n^{th}$ CNT layer can be similarly expressed as described by Eq.(15).

$$r_1 = r_0 + t + x \quad (13)$$

$$r_2 = r_1 + t + x = r_0 + 2t + 2x \quad (14)$$

$$r_n = r_0 + nt + nx \quad (15)$$

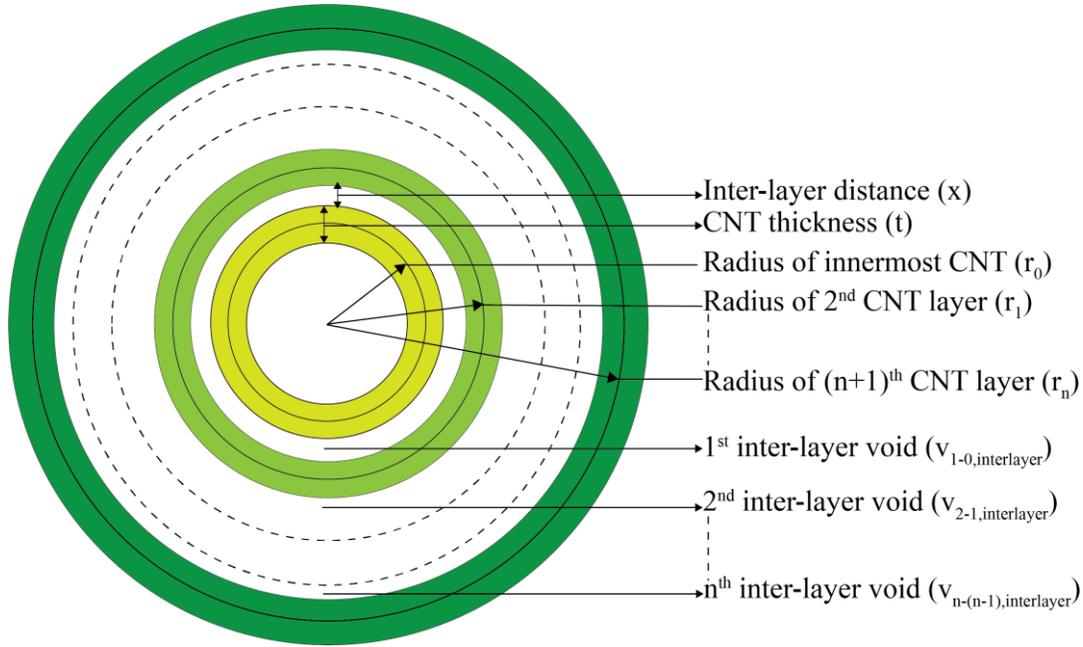

**Figure. 3:** Magnified view of the interlayer distance between the innermost CNT, 2$^{nd,}$ and (n+1)$^{th}$ layer of the MWCNT. The CNT layers between the 2$^{nd,}$ and (n+1)$^{th}$ layer are not shown for clarity.

With the knowledge of the continuum shell model of CNT, the volume of the innermost CNT ($v_{f0}$) can be expressed by Eq.(16), where $l$ denotes the length of the RVE. The subscript '0' of the term $v_{f0}$ denotes the value of '$n$'. The volume of the second layer of the CNT is given by Eq.(17), where the radius of the 2$^{nd}$ layer of the MWCNT is expressed in terms of the $t$, $x$, and $r_0$ using Eq.(15).

$$v_{f0} = \{\pi(r_0 + \frac{t}{2})^2 - \pi\left(r_0 - \frac{t}{2}\right)^2\}l = 2\pi r_0 t l \tag{16}$$

$$v_{f1} = \{\pi(r_1 + \frac{t}{2})^2 - \pi\left(r_1 - \frac{t}{2}\right)^2\}l = 2\pi r_1 t l = 2\pi(r_0 + t + x)t l \tag{17}$$

Similarly, the volume of the third, fourth, and up to (n+1)$^{th}$ layer of the MWCNT can be expressed as below.

$$v_{f2} = \{\pi(r_2 + \frac{t}{2})^2 - \pi\left(r_2 - \frac{t}{2}\right)^2\}l = 2\pi r_2 tl = 2\pi(r_0 + 2t + 2x)tl$$

$$v_{f3} = \{\pi(r_3 + \frac{t}{2})^2 - \pi\left(r_3 - \frac{t}{2}\right)^2\}l = 2\pi r_3 tl = 2\pi(r_0 + 3t + 3x)tl$$

.

.

.

$$v_{fn} = \{\pi(r_n + \frac{t}{2})^2 - \pi\left(r_n - \frac{t}{2}\right)^2\}l = 2\pi r_n tl = 2\pi(r_0 + nt + nx)tl$$

The total volume of the MWCNT is expressed by Eq.(18), after summing and simplifying all the volume terms.

$$v_{f\,total} = 2\pi r_0 tl + 2\pi(r_0 + t + x)tl + 2\pi(r_0 + 2t + 2x)tl + \cdots\cdots 2\pi(r_0 + nt + nx)tl$$

$$\equiv v_{f\,total} = 2\pi r_0 tl + 2\pi tl\{r_0(1 + 1 + \cdots n) + t(1 + 2 + 3 + 4 + \cdots n) + x(1 + 2 + 3 + 4 + \cdots n)\}$$

$$\equiv v_{f\,total} = 2\pi r_0 tl + 2\pi tl\left\{nr_0 + t\frac{n(n+1)}{2} + x\frac{n(n+1)}{2}\right\}$$

$$\equiv v_{f\,total} = 2\pi tl\left\{(n+1)r_0 + t\frac{n(n+1)}{2} + x\frac{n(n+1)}{2}\right\}$$

$$\equiv 2\pi tl(n+1)\left\{r_0 + (t+x)\frac{n}{2}\right\} \quad (18)$$

Using Eq.(18), the expression of the fiber volume fraction ($V_f$) is generated and is shown in Eq.(19). The length of the RVE in lattice units along the x and y-axis is denoted by 'a' (Fig. 4). The lattice parameter of iron is 2.856 Å.

$$V_f = \frac{2\pi tl(n+1)\left\{r_0 + (t+x)\frac{n}{2}\right\}}{a^2(2.856)^2 l} = \frac{2\pi t(n+1)\left\{r_0 + (t+x)\frac{n}{2}\right\}}{a^2(2.856)^2} \quad (19)$$

From Fig. (3) it is evident that there are three types of empty space in the RVE, which might act as voids. The first type of empty space is the interfacial gap between the matrix and the outermost CNT layer, which is considered as a void in the RVE and $v_{v,interface}$ denotes its volume. $v_{v,interface}$ is mathematically modeled as shown in Eq.(20). $h_0$ denotes the distance between the matrix and the midsection of the outermost CNT, which is also the equilibrium distance.

$$v_{v,interface} = \pi\left\{(r_n + h_0)^2 - \left(r_n + \frac{t}{2}\right)^2\right\}l = \pi l\left(h_0 - \frac{t}{2}\right)\left(2r_n + h_0 + \frac{t}{2}\right)$$

$$= \pi l\left(h_0 - \frac{t}{2}\right)\left(2r_0 + 2nx + 2nt + h_0 + \frac{t}{2}\right) \tag{20}$$

The vacant hole inside the innermost CNT is the second type of empty space and is denoted by $v_{v,hole}$. The mathematical expression of this void is shown in Eq.(12).

$$v_{v,hole} = \pi l(r_0 - \frac{t}{2})^2 \tag{21}$$

The third type of empty space is the gap between each CNT layer which is also the graphitic distance [47]. This gap can be considered as a void in the composite or as an interatomic distance. However, the correct assumption is not known. If this inter-layer gap is regarded as void, it can be mathematically modeled as given by Eq.(21). The subscript '1-0' of the term $v_{1-0,interlayer}$ denotes the void between the second MWCNT ($n=1$) layer and the innermost CNT($n=0$).

$$v_{1-0,interlayer} = \pi l\left\{\left(r_1 - \frac{t}{2}\right)^2 - \left(r_0 + \frac{t}{2}\right)^2\right\} = \pi l\{(r_1 + r_0) \ -(r_1 - r_0 - t)\ \}$$

$$= \pi l\{(r_0 + t + x + r_0) \ -(r_0 + t + x - r_0 - t)\ \} = \pi lx(2r_0 + x + t) \tag{22}$$

Similarly, all the interlayer voids in the MWCNT can be expressed as below.

$$v_{2-1,interlayer} = \pi l\left\{\left(r_2 - \frac{t}{2}\right)^2 - \left(r_1 + \frac{t}{2}\right)^2\right\} = \pi l\{(r_2 + r_1) \ -(r_2 - r_1 - t)\ \}$$

$$= \pi l\{(r_0 + 2t + 2x + r_0 + x + t) \ -(r_0 + 2t + 2x - r_0 - t - x - t)\ \}$$

$$= \pi l x (2r_0 + 3x + 3t)$$

$$v_{3-2, interlayer} = \pi l \left\{ \left(r_3 - \frac{t}{2}\right)^2 - \left(r_2 + \frac{t}{2}\right)^2 \right\} = \pi l \{(r_3 + r_2) \ -(r_3 - r_2 - t) \ \}$$

$$= \pi l \{(r_0 + 3t + 3x + r_0 + 2x + 2t) \ - (r_0 + 3t + 3x - r_0 - 2t - 2x - t) \ \}$$

$$= \pi l x (2r_0 + 5x + 5t)$$

.

.

.

$$v_{n-(n-1), interlayer} = \pi l \left\{ \left(r_n - \frac{t}{2}\right)^2 - \left(r_{n-1} + \frac{t}{2}\right)^2 \right\} = \pi l \{(r_n + r_{n-1}) \ -(r_n - r_{n-1} - t) \ \}$$

$$= \pi l \{(r_n + nt + nx + r_0 + (n-1)x + (n-1)t) - (r_0 + nt + nx - r_0 - (n-1)t -$$

$$(n-1)x - t)\}$$

$$= \pi l x (2r_0 + (2n-1)x + (2n-1)t)$$

The sum of all the interlayer void volume is denoted by $v_{tot,interlayer}$ and is given by Eq.(23).

$$v_{tot,interlayer} = \pi l x (2r_0 + x + t) + \pi l x (2r_0 + 3x + 3t) + \pi l x (2r_0 + 5x + 5t) + \cdots \pi l x (2r_0 + (2n-1)x + (2n-1)t)$$

$$\equiv \pi l x \{(2r_0(1 + 1 + 1 + \cdots..) + x(1 + 3 + 5 + \cdots (2n-1)) + t(1 + 3 + 5 \ldots (2n-1)\})$$

$$\equiv \pi l x \{2r_0 n + n^2 x + n^2 t\}$$

$$\equiv \pi l x \{2r_0 n + n^2 (x + t)\}$$

(23)

Summing all three types of voids in the RVE we get the total void volume, which is denoted by $v_{v,tot,1}$ and is shown in Eq.(24).

$$v_{v,tot,1} = v_{v,\text{interface}} + v_{v,\text{hole}} + v_{\text{tot,interlayer}}$$
$$= \pi l(r_0 - \frac{t}{2})^2 + \pi l\left(h_0 - \frac{t}{2}\right)\left(2r_0 + 2nx + 2nt + h_0 + \frac{t}{2}\right) + \pi lx\{2r_0 n + n^2(x+t)\} \quad (24)$$

If the empty spaces between the layers of MWCNT are regarded as interatomic distances, then they are not considered as voids and not included in the total void volume as shown in Eq.(25). $v_{v,tot,2}$ denotes the total void volume when $v_{\text{tot,interlayer}}$ is discarded.

$$v_{v,tot,2} = v_{v,\text{interface}} + v_{v,\text{hole}}$$
$$= \pi l(r_0 - \frac{t}{2})^2 + \pi l\left(h_0 - \frac{t}{2}\right)\left(2r_0 + 2nx + 2nt + h_0 + \frac{t}{2}\right) \quad (25)$$

The void volume fractions ($V_{v,1}$ and $V_{v,2}$) are shown by Eqs. (26) and (27), which were derived by using the two total void volume from Eqs (24) and (25). $V_{v,1}$ denotes the void volume fraction of the RVE when $v_{v,tot,1}$ is considered the total void volume. Similarly, $V_{v,2}$ is the void fraction when $v_{v,tot,2}$ is considered.

$$V_{v,1} = \frac{\pi l(r_0 - \frac{t}{2})^2 + \pi l\left(h_0 - \frac{t}{2}\right)\left(2r_0 + 2nx + 2nt + h_0 + \frac{t}{2}\right) + \pi lx\{2r_0 n + n^2(x+t)\}}{a^2(2.856)^2 l} \quad (26)$$

$$V_{v,2} = \frac{\pi l(r_0 - \frac{t}{2})^2 + \pi l\left(h_0 - \frac{t}{2}\right)\left(2r_0 + 2nx + 2nt + h_0 + \frac{t}{2}\right)}{a^2(2.856)^2 l} \quad (27)$$

From Fig.(2), the matrix volume fraction ($V_m$) can be written in terms of a, $h_0$ and $r_n$ as shown in Eq.(28). Since the matrix volume fraction is not changing, it is considered to be equal to a constant '$K$'. Eq.(28), can be used to express the lattice parameter 'a' in terms of $n$ and other constants as described by Eq.(29).

$$V_m = K = 1 - \frac{\pi(r_n + h_0)^2}{a^2 2.856^2} \tag{28}$$

$$\equiv \frac{\pi(r_n + h_0)^2}{a^2 2.856^2} = (1 - K)$$

$$\equiv a^2 = \frac{\pi(r_n + h_0)^2}{(1 - K) \, 2.856^2}$$

$$\equiv a^2 = \frac{\pi(r_0 + nt + nx + h_0)^2}{(1 - K) \, 2.856^2} \tag{29}$$

Using the value of $a^2$ from Eq.(29) in Eqs. (19), (26) and (27); $V_f, V_{v,1}$ and $V_{v,2}$ can be expressed in terms of $n$ and other constants as described by Eqs. (30), (31), and (32).

$$V_f = \frac{2t(1-K)(n+1)\left\{r_0 + (t+x)\frac{n}{2}\right\}}{(r_0 + nt + nx + h_0)^2} \tag{30}$$

$$V_{v,1} = \frac{(1-K)\left[(r_0 - \frac{t}{2})^2 + \left(h_0 - \frac{t}{2}\right)\left(2r_0 + 2nx + 2nt + h_0 + \frac{t}{2}\right) + x\{2r_0 n + n^2(x+t)\}\right]}{(r_0 + nt + nx + h_0)^2} \tag{31}$$

$$V_{v,2} = \frac{(1-K)\left[(r_0 - \frac{t}{2})^2 + \left(h_0 - \frac{t}{2}\right)\left(2r_0 + 2nx + 2nt + h_0 + \frac{t}{2}\right)\right]}{(r_0 + nt + nx + h_0)^2} \tag{32}$$

Expressions of $V_f, V_{v,1}$ and $V_{v,2}$ from Eqs. (30), (31) and (32) are used to develop two expressions of the function $Z$ as shown in Eqs (33) and (34). The difference between $Z_1$ and $Z_2$ lies in the consideration of the expression of void volume fraction ($V_v$). In $Z_1$ the void volume fraction is $V_{v,1}$ which considers the interlayer gaps between layers of the MWCNT as voids. Conversely, $Z_2$ regards $V_{v,2}$ as the total void fraction which does not consider the interlayer gap as void.

$$Z_1 = V_f - V_{v,1} =$$

$$\frac{2t(1-K)(n+1)\left\{r_0 + (t+x)\frac{n}{2}\right\}}{(r_0 + nt + nx + h_0)^2} - \frac{(1-K)\left[(r_0 - \frac{t}{2})^2 + \left(h_0 - \frac{t}{2}\right)\left(2r_0 + 2nx + 2nt + h_0 + \frac{t}{2}\right) + x\{2r_0 n + n^2(x+t)\}\right]}{(r_0 + nt + nx + h_0)^2}$$

$$\tag{33}$$

$$Z_2 = V_f - V_{v,2} =$$

$$\frac{2t(1-K)(n+1)\left\{r_0 + (t+x)\frac{n}{2}\right\}}{(r_0 + nt + nx + h_0)^2} - \frac{(1-K)[(r_0 - \frac{t}{2})^2 + (h_0 - \frac{t}{2})(2r_0 + 2nx + 2nt + h_0 + \frac{t}{2})]}{(r_0 + nt + nx + h_0)^2}$$

(34)

The values of $E_c$ and $\sigma_{c,ut}$ of RVE containing two ($n$=1), four ($n$=3), eight ($n$=7), and thirteen ($n$ =12) layered MWCNT were found by performing uniaxial tension simulation.

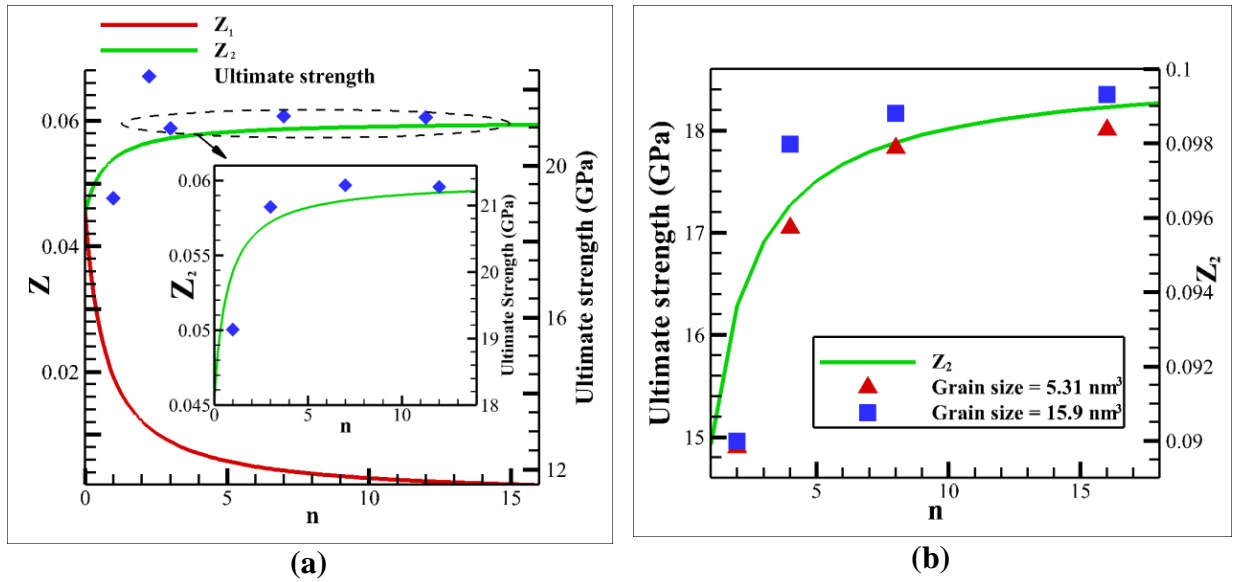

(a)            (b)

**Figure. 4:** **(a)** variation of $\sigma_{c,ut}$(ultimate strength), $Z_1$ and $Z_2$ with $n$ (CNT layer number). The values of $\sigma_{c,ut}$ are obtained from simulation. It also shows a magnified view of $Z_2$ and $\sigma_{c,ut}$ vs $n$. **(b)** The variation of $\sigma_{c,ut}$ (obtained from the simulation) and $Z_2$ with $n$ (CNT layer number) for RVEs respectively having an average grain size of 5.31 nm³ and 15.9 nm³.

From Fig.4(a), it is evident that $Z_2$ correctly predicts the pattern of variation of $\sigma_{c,ut}$ with $n$. Similarly, it was observed that $E_c$ also changes with $n$ according to the shape of $Z_2$. Therefore, it can be concluded that the interlayer gap or the graphitic distance doesn't act as a void in the composite and doesn't affect the composite's strength and stiffness. From the magnified view of $Z_2$ in Fig.4(a) it can be observed that $Z_2$ changes drastically from $n$=0 to $n$=3, which suggests a high change in composite's strength in that range. The values of $\sigma_{c,ut}$ obtained from the simulation are also in accordance with this prediction. The $\sigma_{c,ut}$ of composites reinforced with four-layered MWCNT are 9.8% greater than the ones reinforced with double-layered MWCNT. For $n=4$ and greater, $Z_2$ increases slightly, which indicates a small change in strength between composites strengthened with CNT having five or more layers. Results from the simulation also verify this prediction as to the average increase of $\sigma_{c,ut}$ is around 2.1% between the RVE reinforced with four, eight, and thirteen layers. Similarly from the data obtained from the simulations, it was observed that Young's modulus increases by 7.2% between RVE separately having double and four-layered MWCNT but the improvement in $E_c$ for increasing the reinforcing MWCNT layer number from four to eight and eight to thirteen are 1.35% and 0.33% respectively, which is quite negligible.

From the above discussions, it is evident that using MWCNT having less than four layers results in weaker CNT-Fe composites but increasing the layer number more than four does not result in significant improvement of mechanical properties. Moreover, the mechanical properties between composites synthesized separately with two, three, four, and five-layered MWCNT vary greatly.

When manufacturing composites the matrix is constituted of several grains instead of a single crystal. When the number of layers of the reinforcing MWCNT increases, its overall diameter also increases, which could result in the interface coming in contact with a large number of grain boundaries. The grain boundaries can interact with the interface and have a detrimental compounding effect in reducing the strength of the composite and lead the pattern of strength and stiffness away from the predicted model. Therefore, it was necessary to validate the predicted trend for composites reinforced with different layers of MWCNT and having different grain sizes. From Fig.4(b) it is evident that the strength and the stiffness of the composite decreases with the decreasing grain sizes but for a fixed average grain size the overall pattern of the variation of $\sigma_{c,ut}$ is similar to $Z_2$. For increasing the layer number of the reinforcing MWCNT from two to four, from four to eight, and finally from eight to sixteen the composite's strength increases 14.43%, 4.55%, 1.02% and 18.89%, 3.1%, 1.12% for an average grain size of 5.31 nm$^3$ and 15.9 nm$^3$ respectively, which is very close to the prediction of $Z_2$. Therefore it can be concluded that the theoretical model holds even for multi-grained composites and the failure depends less on the interface between the CNT and matrix.

## 3.2 FAILURE MECHANISM AND THE EFFECT OF GRAIN SIZE:

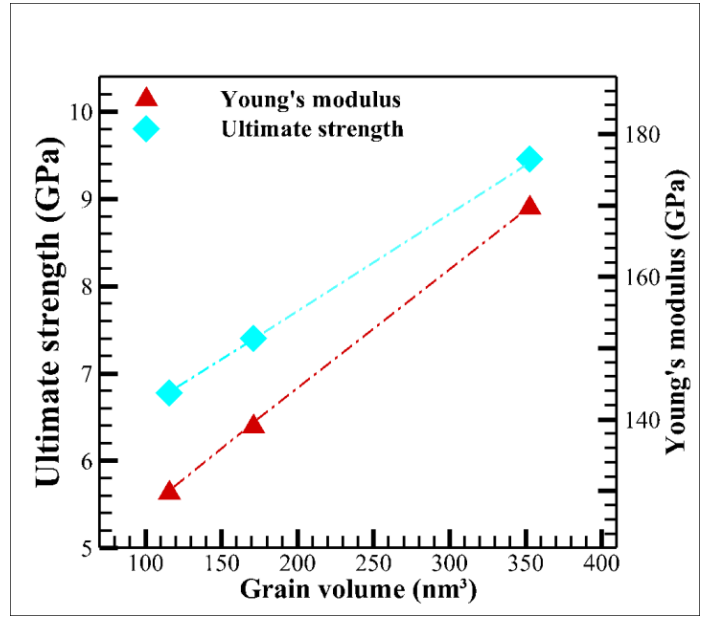

**Figure. 5:** Variation of ultimate strength and young's modulus with average grain volume.

The grain size of the matrix affects the mechanical properties of the composite to a great extent. For investigating the effect of grain size, uniaxial tension simulation was performed on three RVE models of CNT-Fe composite, which are reinforced by a (39,39) SWCNT and has an average grain size of 352.88 nm$^3$, 171.12 nm$^3$ and 115.67 nm$^3$ respectively. SWCNT was used in these models because we were only focusing on grain behavior and having extra atoms inside the CNT will increase the unnecessary computational load. From Fig.(5) it is evident that even for with large RVE models, the strength and stiffness linearly decrease with the reduction of average grain size. For every 50% decrease in grain size, the ultimate strength and stiffness decrease 28.02% and 18.03% respectively. This phenomenon is known as the 'inverse hall petch effect' [48], [49] and it arises when the grain sizes of metals are below the critical grain size (10 nm) [50]. Therefore, it can also be concluded that embedding CNT does not drastically change the overall tendency of the composite to follow its matrix material.

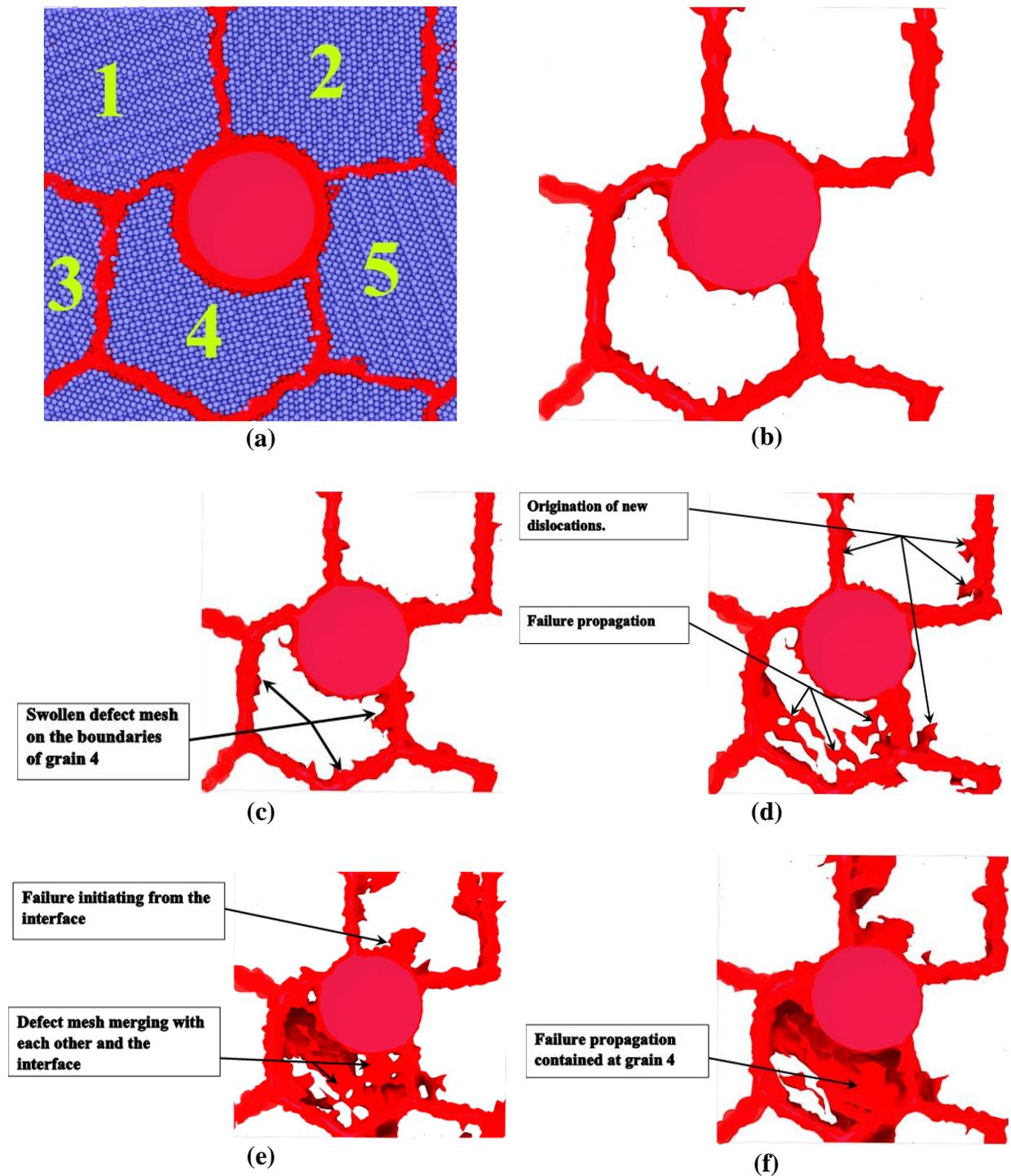

**Figure. 6:** These pictures show the development of failure by DXA analysis of RVE having an average grain volume of 352.88 nm$^3$. **(a)** Shows the top view of the RVE before applying stress. The grains are numbered for identification. **(b)** shows the same image without the atoms and grain numbers. **(c)**,**(d)**,**(e)** and **(f)** show the condition of the defect mesh at strain 0.063, 0.068, 0.072 and 0.076 respectively.

The failure mechanism of the poly-nanograined composite is still a mystery because most of the MD studies were performed on single-crystal composite [43]. Since, all RVE models having different grain sizes shows similar failure mechanism, only the failure process of the RVE having an average grain volume of 352.88 nm$^3$ is explained. The grains relevant for this analysis are numbered for identification as shown in Fig. 6(a). The dislocation extraction algorithm (DXA) was used in this study to locate the initiation and propagation of dislocation. In this method, a red mesh covers the atoms that do not have an ideal bcc configuration. As a result, the atoms near the grain boundaries, interface, and the CNT are covered in red mesh from the beginning of the simulation as shown in Fig.6(b). For visual clarity, the atoms are hidden and only the red mesh is made visible from Figs. 6(c) to 6(f). The failure of the composite begins at strain 0.063. At this strain, the defect mesh covering the boundary of grain 4 starts to swell and increase in size which suggests the displacement of atoms from their lattice positions or the initiation of dislocation. At 0.068 strain, the dislocations propagate within grain 4 and new dislocation starts to spread in grain 2 and 5 from their boundaries. When the composite is near strain 0.072, the dislocations inside grain 4 merge with the CNT-Fe interface, which is revealed because of the connection of the defect meshes (Fig.6(e)). At this strain, dislocations also start to emerge from the interface, which can be seen in grain 2 of Fig.6(e). Therefore, it can be concluded that for nano-grained CNT-Fe composite the failure is initiated from the grain boundary. This finding is highly in contrast with the results obtained from previous MD studies that considered single-crystal RVE, which reported that failure originates from the matrix-fiber interface. Since real materials have grains, the failure process observed in this study should be more accurate. Moreover, our analysis also reveals that dislocation is the primary reason for inverse hall petch behavior (Fig.6), which is reported from other studies as well [51]. From Fig.6(f), it can be observed that the defect mesh is constrained within the boundaries of grain 4. This is known as a dislocation pile-up [52], in which the grain boundaries restrict the propagation of dislocation. Due to this phenomenon, nano grained CNT-Fe fails slowly and in a longer strain range than single-crystal CNT-Fe composites.

# 4. CONCLUSION:

In this study, we performed MD simulations and mathematical modeling to gain an understanding of factors affecting the mechanical properties of CNT-Fe composites. A novel mathematical model was developed and verified by MD simulation, which revealed that the strength and stiffness of CNT-Fe composites reinforced with MWCNTs increase significantly when the number of layers of the MWCNT is varied from one to eight. It was also discovered that for composites reinforced with MWCNTs the empty spaces present between the CNT layers do not act as a void and have no effect on its mechanical properties. The theoretical model was validated for multigrain matrices. It was found that, in nano-grained CNT-Fe composites, failure develops from the grain boundaries of the matrix and dislocation is the primary mechanism of the observed inverse hall petch effect. This study enhances the present understanding and provides novel insights for manufacturers which will be extremely beneficial for economically synthesizing the composites for multipurpose applications and also for improving its mechanical properties. The developed theoretical model can be used for any metal matrix composites by appropriately changing the parameters. The method for theoretically modeling RVE to determine the change of the desired properties of the composite can also be used to inspect many other factors affecting the composite, such as hardness, coefficient of thermal expansion, crystal defects and also the reinforcing effect of different types of fibers.

## Data availability

The raw data required to reproduce these findings cannot be shared at this time due to legal or ethical reasons. However, the data can be made available on request from the corresponding author, upon reasonable request.

# Acknowledgments

The authors of this paper would like to express their gratitude to the Department of Mechanical Engineering, BUET and Multiscale Mechanical Modelling and Research Network (MMMRN) group for providing the computational facilities and technical support.